\documentclass[sigplan, noacm]{acmart}

\usepackage{algorithm,algorithmicx}
\usepackage[noend]{algpseudocode}
\usepackage{amsmath,amssymb,amsfonts}
\usepackage[title]{appendix}
\usepackage{array}
\usepackage{booktabs}
\usepackage{caption,subcaption}
\usepackage{circledsteps}
\usepackage{color}
\usepackage[inline]{enumitem}
\usepackage{filecontents}
\usepackage{graphicx}
\usepackage{multirow}
\usepackage{pifont}
\usepackage{siunitx}
\usepackage{tabularx}
\usepackage{tikz}
\usepackage{titlesec}
\usepackage{ulem}
\usepackage{xspace}
\usepackage{xurl}

\iffalse

\else

\fi

\renewcommand{\algorithmicrequire}{\textbf{Input:}}
\renewcommand{\algorithmicensure}{\textbf{Output:}}

\newcommand{\systemname}{\textsc{AverSearch}\xspace}
\newcolumntype{x}[1]{>{\centering\arraybackslash}p{#1}}

\def\BibTeX{{\rm B\kern-.05em{\sc i\kern-.025em b}\kern-.08em
    T\kern-.1667em\lower.7ex\hbox{E}\kern-.125emX}}

\begin{document}

\title{Efficient Graph-Based Approximate Nearest Neighbor Search Achieving: Low Latency Without Throughput Loss}


\author{Jingjia Luo, Mingxing Zhang, Kang Chen \\
Xia Liao, Yingdi Shan, Jinlei Jiang, Yongwei Wu}

\affiliation{%
  \institution{Tsinghua University}
  \city{Beijing}
  \country{China}
}

\email{
  luojj22@mails.tsinghua.edu.cn,
  zhang\_mingxing@mail.tsinghua.edu.cn,
  chenkang@tsinghua.edu.cn,}
\email{liaoxia5018@163.com,
  shanyd@tsinghua.edu.cn,
  jjlei@tsinghua.edu.cn,
  wuyw@tsinghua.edu.cn}

\begin{abstract}
The increase in the dimensionality of neural embedding models has enhanced the accuracy of semantic search capabilities but also amplified the computational demands for Approximate Nearest Neighbor Searches (ANNS).
This complexity poses significant challenges in online and interactive services, where query latency is a critical performance metric.
Traditional graph-based ANNS methods, while effective for managing large datasets, often experience substantial throughput reductions when scaled for intra-query parallelism to minimize latency.
This reduction is largely due to inherent inefficiencies in the conventional fork-join parallelism model.

To address this problem, we introduce \systemname, a novel parallel graph-based ANNS framework that overcomes these limitations through a fully asynchronous architecture.
Unlike existing frameworks that struggle with balancing latency and throughput, \systemname utilizes a dynamic workload balancing mechanism that supports continuous, dependency-free processing.
This approach not only minimizes latency by eliminating unnecessary synchronization and redundant vertex processing but also maintains high throughput levels.
Our evaluations across various datasets, including both traditional benchmarks and modern large-scale model generated datasets, show that \systemname consistently outperforms current state-of-the-art systems.
It achieves up to 2.1$\times$-8.9$\times$ higher throughput at comparable latency levels across different datasets and reduces minimum latency by 1.5$\times$ to 1.9$\times$.
\end{abstract}

\maketitle

\section{Introduction}

Scaling laws for large language models indicate that increasing model size enhances performance. 
As a result, the dimensionality of embedding vectors for objects like documents and images has expanded from approximately 100 dimensions in earlier benchmarks (e.g., SIFT~\cite{sift1b}, DEEP~\cite{deep1b}) to thousands in recent models (e.g., OpenAI  ~\cite{wikipedia-supabase}, Cohere~\cite{cohere2022multilingual}). This growth not only improves the accuracy of similarity measurements but also significantly elevates the computational demands of conducting Approximate Nearest Neighbor Searches (ANNS). 

For instance, inspired by OpenAI o1's advancements~\cite{openai2024o1}, Retrieval-Augmented Generation (RAG)~\cite{yue2024inference-scaling} is utilized to improve the generation quality, which
increases with the number of retrieval rounds, causing the response time to scale proportionally with retrieval latency.
Additionally, in attempts using ANNS in sparse attention calculations~\cite{liu2024retrievalattention, chen2024magicpig, zhang2024pqcache}, retrieval occurs for every layer and token, further emphasizing the need to minimize latency. 
This scenario presents a challenging balancing act among {\bf accuracy, throughput, and latency}, a trio of metrics that often conflict with one another.
In scenarios with stringent low-latency requirements, \textbf{goodput}, i.e., throughput that meets specific latency constraints, must also be considered.

Researchers have developed a multitude of algorithms to improve searching performance, including hashing-based~\cite{andoni2015practical, lu2020vhp, chen2021spann}, quantization-based~\cite{wei2020analyticdb, wu2017multiscale, ge2013optimized}, tree and graph based methods~\cite{silpa2008optimised, wang2020deltapq, muja2009fast, fu2016efanna, fu2017fast, fu2021high, malkov2014approximate, malkov2018efficient, ren2020hm, munoz2019hierarchical, zhao2023towards}, and the combinations of multiple methods~\cite{jayaram2019diskann, chen2018sptag, yang2024fast, lu2024probabilistic}. 
Graph-based algorithms, in particular, are proven by many studies~\cite{web-annbench, aumuller2020annbench, web-bigannbench} that excel at delivering optimal throughput-recall trade-offs.
Many benchmark studies~\cite{web-annbench, web-bigannbench, aumuller2020annbench} have shown that graph-based algorithms excel at delivering optimal throughput-recall trade-offs.

\begin{figure}[ht]
  \centering
  \includegraphics[width=0.9\linewidth]{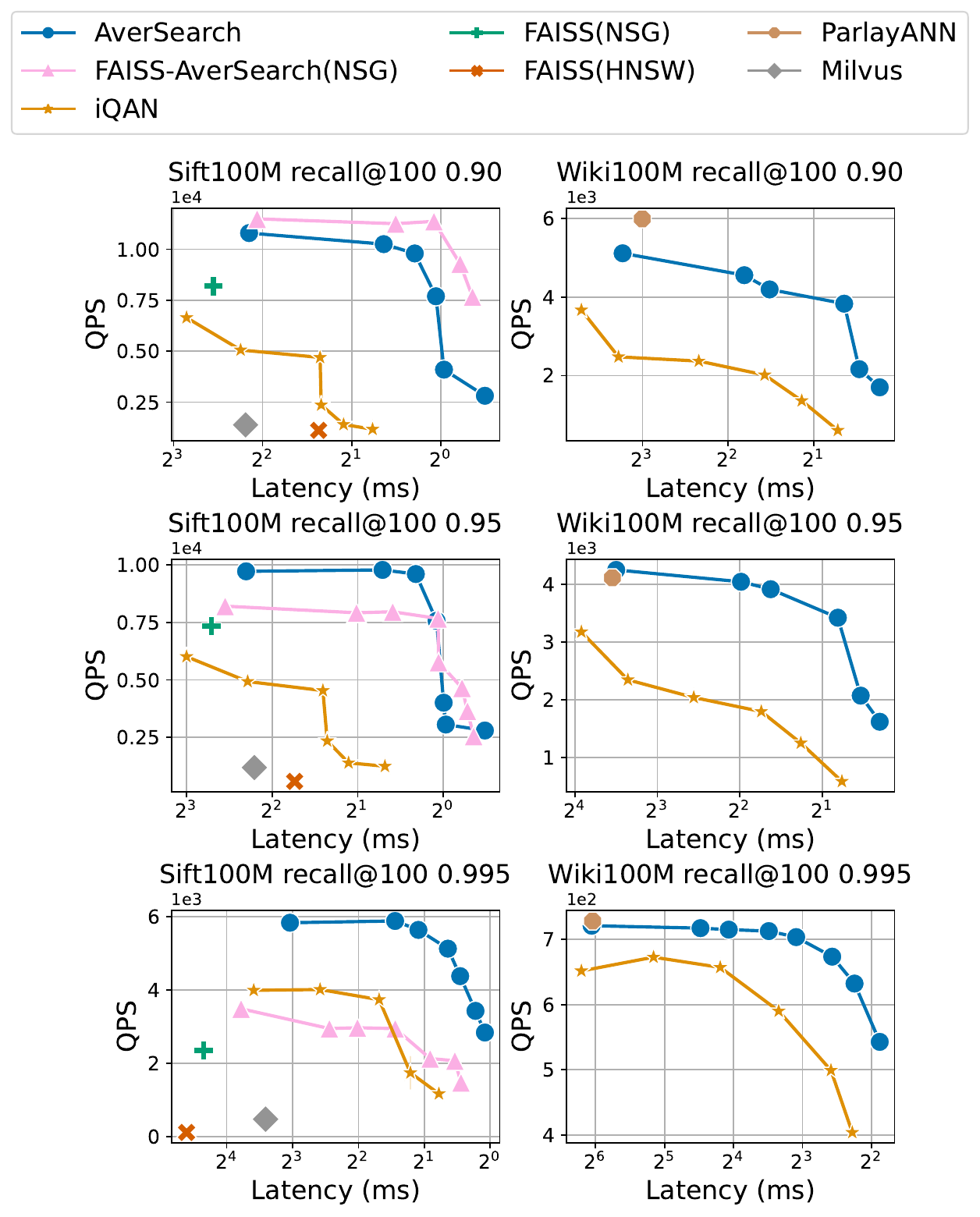}
  \caption{The relationship between latency and throughput, expressed in queries per second (QPS), varies across different parallelism settings.
  We leverage all 48 available cores, organizing them into ``$intra \times inter$'' groups, ranging from ``1 $\times$ 48'' to  ``24 $\times$ 2''.
  Here, ``$intra$'' represents the number of threads dedicated to each query, while ``$inter$'' indicates the number of independent concurrent queries.
  The analysis uses two well-known datasets, SIFT100M and Wiki100M~\cite{wikipedia-cohere-2022}, each containing 100 million vectors with dimensions of 128 and 768, respectively. 
  The evaluation covers various recall levels from 0.9 to 0.995.
  }
  \label{fig:eval-deep-sift}
\end{figure}

\begin{figure}[b]
  \centering
  \begin{subfigure}[b]{0.47\linewidth}
    \centering
    \centerline{\includegraphics[width=\linewidth]{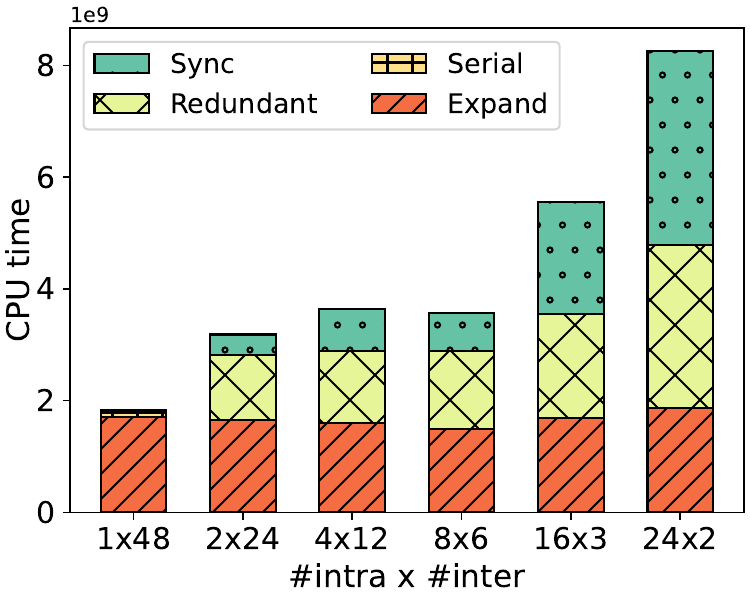}}
    \caption{iQAN CPU time breakdown over a query}
    \label{fig:CPU-time-simple}
  \end{subfigure}
  \hfill
  \begin{subfigure}[b]{0.46\linewidth}
    \centering
    \centerline{\includegraphics[width=\linewidth]{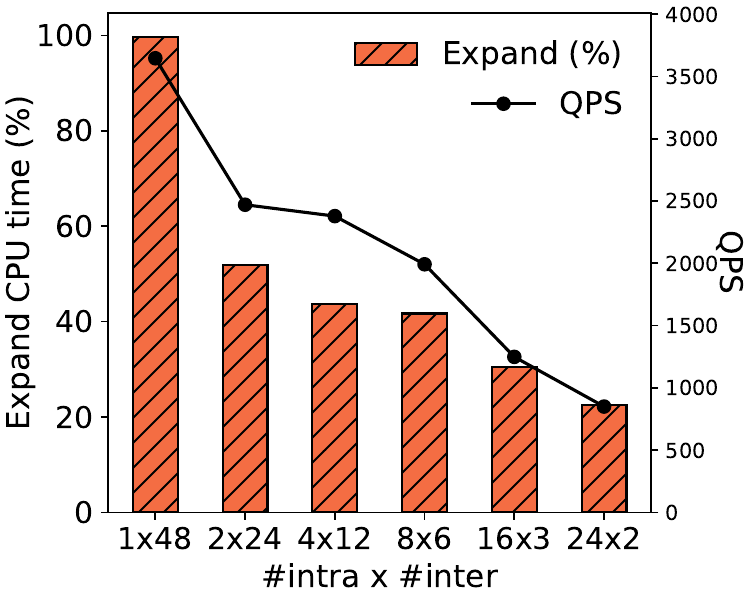}}
    \caption{iQAN valid expansion CPU time percentage V.S. QPS}
    \label{fig:CPU-time-perc-simple}
  \end{subfigure}
  \caption{iQAN execution time distribution under different parallelism strategies with Wiki100M and recall at 0.9.}
\end{figure}

Traditional approaches enhance search performance on graphs by optimizing their structures, whereas recent studies~\cite{yangcspg, peng2023iqan} incorporate intra-query parallelism to reduce latency.
However, implementing such parallel strategies presents significant challenges.
As an illustration, Figure \ref{fig:eval-deep-sift} presents the throughput-latency trade-off across different recall levels using iQAN~\cite{peng2023iqan}, a state-of-the-art parallel graph-based ANNS engine. 
The figure shows that while increasing intra-query parallelism decreases latency, it also leads to a drop in overall throughput.
In the Sift100M 0.995 benchmark, the QPS drops from 3993 to 1166 as the number of intra-query threads increases from 1 to 24, utilizing all 48 threads.

Further analysis reveals that this problem stems from {\bf inherent inefficiencies in the traditional fork-join model} of parallelism. 
Even with a balanced workload distribution, variations in execution times among threads cause substantial synchronization overhead. 
Figure \ref{fig:CPU-time-simple} breaks down the execution time under different intra-query parallelism settings, highlighting that an increase in intra-query parallelism exacerbates synchronization and redundant processing burdens.
This escalation reduces the proportion of genuinely effective work (the ``expand'' portion), directly contributing to a decline in overall throughput, as depicted in Figure \ref{fig:CPU-time-perc-simple}. 
More details on this analysis will be provided in \S\ref{sec:insight}.

In response to these challenges, we introduce \systemname (\textbf{A}synchronous \textbf{ver}tex \textbf{Search}),
a parallel graph-based ANNS framework that maintains high throughput while simultaneously reducing latency through enhanced intra-query parallelism. 
The cornerstone of \systemname is its innovative {\bf fully asynchronous architecture}, which divides the job of query processing among three distinct roles of threads, that crucially, rarely wait on each other. 
This architecture allows memory-intensive distance calculation to continuously proceed without delays, thereby optimizing memory bandwidth utilization.
Moreover, \systemname utilizes a dynamic online workload balancing approach, enabling faster threads to speculatively process additional vertices rather than idling, while slower threads move on to subsequent iterations without being burdened with unnecessary vertices.
This strategy leads to near-zero synchronization overhead and minimizes the processing of redundant tasks.

To evaluate the effectiveness of \systemname, we conducted experiments on various datasets, including classical ones with approximately 100-dimensional vectors as well as recent large model-generated datasets from OpenAI~\cite{wikipedia-supabase, wikipedia-stephanst} and Cohere~\cite{wikipedia-cohere-2022} (with dimensions exceeding a thousand).
We compared \systemname not only with iQAN, the leading parallel ANNS engine but also with prominent commercial vector search engines like Milvus~\cite{wang2021milvus, guo2022manu} and FAISS~\cite{douze2024faiss} across various settings. 
Results demonstrate that \systemname achieves up to 2.1$\times$-8.9$\times$ higher throughput at comparable latency levels across different datasets.
Moreover, it is capable of achieving 1.5$\times$ to 1.9$\times$ lower minimum average latency due to better intra-query parallelism utilization.
Detailed analysis involving memory utilization, vertex processing redundancy, and comparing with quantization-based indexes further justifies our observations and pinpoints the sources of optimization.
These findings highlight the capacity of \systemname to effectively manage the complex trade-offs between throughput, accuracy, and latency in real-time search applications.

\section{Background and Related Works}
\label{sec:background}

\subsection{Approximate Nearest Neighbor Search}

In high-dimensional embedding spaces, similarity between entities is measured by metrics like inner product or the Euclidean norm, which supports vector search techniques for identifying the top $K$ nearest vectors to a query.
The approximate version of this process, known as ANNS, aims to efficiently find the approximate $K$ closest vectors to a query vector $q \in \mathbb{R}^d$ within a large set of $N$ vectors in a $d$-dimensional space.
The effectiveness of ANNS is measured by its ability to include the actual nearest neighbors among the $K$ vectors identified.

Various indexing ANNS strategies to prevent full database scan include tree-based~\cite{silpa2008optimised, wang2020deltapq, yesantharao2022parallel}, hash-based~\cite{andoni2015practical, lu2020vhp, sundaram2013streaming}, quantization-based~\cite{wei2020analyticdb, wu2017multiscale}, and graph-based methods. 
These non-graph-based methods rely on partitioning and clustering the vector database, using pretrained data structures, hashing functions, or quantization codes to enable quick identification of clusters nearest to a query vector.
However, they face challenges with scalability and efficiency in high-dimensional spaces due to the well-known ``curse of dimensionality''. 
Prior works on graph-based indices have demonstrated superior performance compared to state-of-the-art non-graph-based methods, achieving QPS improvements of up to 20 times~\cite{fu2017fast, fu2021high, malkov2018efficient, harwood2016fanng, hajebi2011fast}.
Additionally, graph-based methods~\cite{web-ngt, web-qsgngt} maintain a leading position in achieving a higher QPS-recall trade-off in the ANN Benchmarks~\cite{aumuller2020annbench, web-annbench}.

\subsection{The Latency Requirements for ANNS}

With the increasing dimensionality and the adoption of RAG in online and interactive services, latency requirements are becoming more stringent, necessitating a focus beyond throughput alone.
Firstly, Large embedding models significantly increase the dimensionality of embedding vectors, thereby introducing substantial latency overhead.
Moreover, following the inference-time scaling law, a single user request may demand multiple rounds of RAG to generate a response.
Consequently, response time scales proportionally with retrieval latency multiplied by the number of retrievals.
Lastly, ANNS is also increasingly applied in long-context LLM inference for attention retrieval~\cite{liu2024retrievalattention, chen2024magicpig, zhang2024pqcache}.
Due to the sparsity of attention, the algorithms leverage softmax's natural ability to select key-value pairs with high attention scores, 
where retrieval occurs for every layer and token in a serial manner. 
Therefore, ANNS systems should adopt new design objectives to accommodate emerging application scenarios.
Under strict latency constraints, considering goodput-throughput under varying latency conditions—can better address these evolving demands.

\subsection{Graph Index Searching and Existing Parallelizing Method}

\subsubsection{Best-First Search algorithm (BFiS)}
\label{sec:bfis}

\begin{figure*}[h]
  \begin{subfigure}[c]{0.17\linewidth}
    \centering
    \centerline{
      \includegraphics[width=\linewidth]{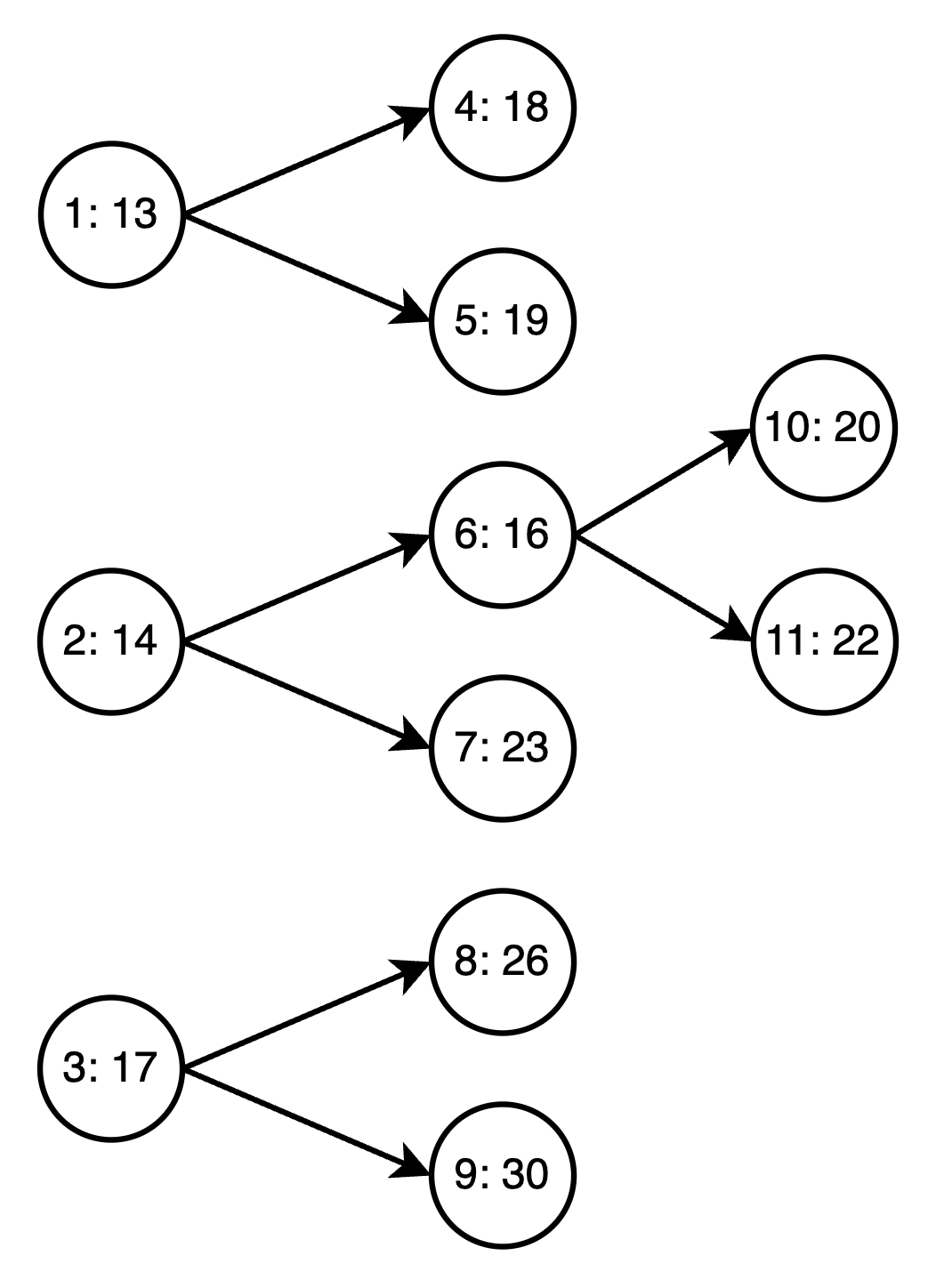}}
    \caption{
      A directed similarity graph.}
    \label{fig:example-graph-0}
  \end{subfigure}
  \hfill
  \begin{subfigure}[c]{0.82\linewidth}
    \centering
    \centerline{
      \includegraphics[width=\linewidth]{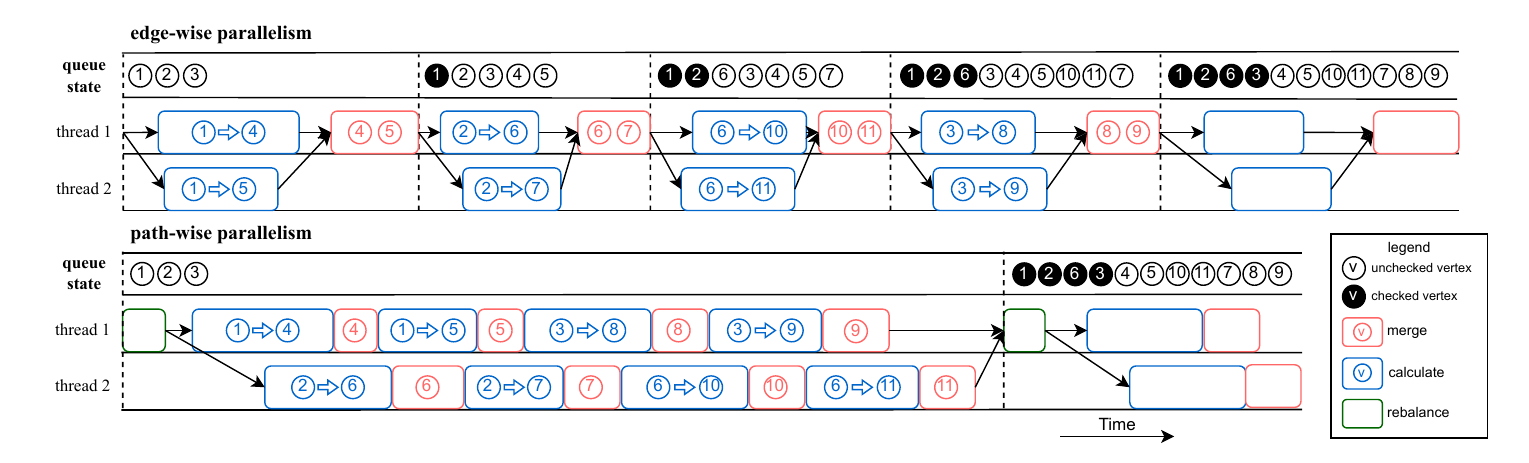}}
    \caption{
      A timeline depicting different parallelism strategies.
      In the figure, ``queue state'' represents the evolving state of the global priority queue over time. "$\textcircled{x} \Rightarrow \textcircled{y}$" indicates edge $(x, y)$ in the similarity graph.}
    \label{fig:edge-wise-vs-path-wise}
  \end{subfigure}
  \caption{Edge-wise parallelism v.s. path-wise parallelism (\S\ref{sec:pbfis})}
  \label{fig:pbfis}
\end{figure*}

In graph-based ANNS, a similarity graph $G(V, E)$ is constructed, where each vertex $v \in V$ represents a vector, and each edge $(v, u) \in E$ indicates that $u$ is among the closest neighbors of $v$.
Despite the variety of methods used to construct $G$, a unified Best-First Search algorithm (BFiS) is applied across all graph-based indices for sequential querying. 
BFiS initiates its search from a seed vertex, selected either randomly or based on the graph's specifics, and progresses toward the query by navigating through outgoing edges.  
As we can see from Algorithm \ref{algo:BFiS}, it utilizes a priority queue $Q$ to maintain focus solely on the $L$ unchecked nodes nearest to the query (l\ref{alg:resize}), 
with the $L$ parameter being crucial for controling the search precision. 
In each iteration, BFiS identifies the nearest unchecked node, and proceeds to expand it.
To facilitate further discussion, we introduce the term \textbf{expand} as calculating the pairwise distance between all neighbors of the nearest unchecked node and the query (l\ref{alg:candi}), marking it as checked, and subsequently adding its neighbors to the priority queue as new unchecked candidates for future expansion. 
The queue is sorted by distance to the query, ensuring that less promising candidates are replaced as new ones are introduced.
The search expands unchecked nodes until none remain in the priority queue.


\begin{algorithm}
  \caption{Best-First Search \label{algo:BFiS}}
  \label{alg:bfis}
  \begin{algorithmic}[1]
    \Require {
        query $q$,
        priority queue $Q$ initialized with entry-node set,
        graph $G$,
        distance function $\delta$,
        queue capacity $L$,
        number of ANNs $K$}
    \Ensure {$K$ approximate nearest neighbors of $q$}
    \While {there's unchecked candidate in $Q$}
      \State {\color{blue} \# path-wise parallel knob in iQAN \{} \label{alg:path-wise}
        \State \quad $v \gets$ the first unchecked candidate in $Q$ \label{alg:candi}
        \State \quad $Expand(v, q, Q, G, \delta)$
      \State {\color{blue} \}}
      \If {$|Q| > L$} 
        \State {$Q.resize(L)$} \label{alg:resize}
      \EndIf
    \EndWhile
    \Return the first $K$ vertices in $Q$
  \end{algorithmic}
\end{algorithm}

\begin{algorithm}
  \caption{the Expand operation}
  \label{alg:expand}
  \begin{algorithmic}[1]
    \Require {
        the expanded candidate $v$,
        query $q$,
        priority queue $Q$,
        graph $G$,
        distance function $\delta$
    }
    \State {$v.state \gets checked$}
    \State {$B \gets \emptyset$}
    \State {\color{red} \# edge-wise parallel knob} \label{alg:edge-wise}
    \ForAll {neighbor $u$ of $v$ in $G$}
        \If {$u$ is not visited}
          \State {mark $u$ as visited}
          \State {$u.state \gets unchecked$}
          \State {$u.distance \gets \delta(u, q)$}
          \State {$B \gets B \cup u$}
          \label{alg:calculation}
        \EndIf
    \EndFor
    \State $Q \gets Q.merge(B)$ \label{alg:reduce}
  \end{algorithmic}
\end{algorithm}

\subsubsection{Parallel BFiS}
\label{sec:pbfis}

Distributing different queries across threads sufficed to achieve high throughput but leads to significant latency, especially with large and high-dimensional databases.
For example, searching a Wiki100M dataset on a single CPU thread could take up to 73.774 ms for each query, a duration that is impractical for real-time applications like retrieval augmented generation (RAG) and the whole Wikipedia contain even much more pages than 100M.
Consequently, leveraging intra-query parallelism becomes imperative for latency reduction.

Traditional graph processing systems~\cite{nguyen2013lightweight, chi2016nxgraph, han2015giraph} support parallel BFS, but they struggle with BFiS due to strict data dependencies and frequent synchronizations thus a specialized solution is needed.
A direct strategy for intra-query parallelism is \textit{edge-wise} parallelism, which delegates distance calculation for different neighbors of the active vertex to various threads (i.e. parallelizing the for loop bellowing l\ref{alg:edge-wise} of Algorithm \ref{alg:expand}).
However, this approach is hampered by the considerable overhead from frequent synchronization.
To circumvent this, the state-of-the-art parallel graph index ANNS algorithm, iQAN \cite{peng2023iqan}, observes that candidates at the forefront of $Q$ are highly likely to be expanded in subsequent steps, permitting the speculative expansion of multiple candidates without significantly compromising accuracy.

Therefore, iQAN adopts \textit{path-wise} parallelism.
Essentially, it parallelizes the code block below l\ref{alg:path-wise} of Algorithm \ref{algo:BFiS}.
With a larger parallel section, iQAN significantly reduces the synchronization frequency.
It distributes the priority queue into thread-local queues before each parallel epoch, allowing each thread to expand its assigned paths independently.
However, global synchronization is still necessary for global queue pruning and rebalancing among thread-local queues.

\subsubsection{Example}
\label{sec:example}

Figure \ref{fig:pbfis} compares edge-wise and path-wise parallelism with an example graph, where vertices are represented by circles and edges depicted by arrows.
In the graph each vertex is labeled in the in the format ``ID: distance'' to denote its ID and the distance to the query.
As we can see, with edge-wise parallelism, the two outgoing edges of vertex \ding{192} are assigned to different threads, which calculate the distance between the query and vertex \ding{195} and \ding{196} independently.
This parallel calculation is immediately subsequent to a global synchronization for merging \ding{195} and \ding{196} into the priority queue. 

In contrast, with a predefined $width$ of 2 in path-wise parallelism, 
each of the two threads can expand 2 candidates independently in the subsequent parallel epoch.
During the parallel epoch, each thread consistently expands the first unchecked vertex in its local sub-queue.
This process repeats until all vertices in the sub-queue are checked or the preset expansion limit (2 in this case) is reached.

As depicted in Figure \ref{fig:pbfis} (b), the three unchecked vertices in the global queue are initially distributed to each thread's local sub-queues in a round-robin fashion.
Thread 1 then expands vertices \ding{192} and \ding{194}, incorporating their neighbors into its local sub-queue.
Similarly, Thread 2 expands \ding{193}, 
and subsequently expand \ding{197} without waiting for global synchronization.

However, iQAN's path-wise parallelism reduces the frequency of global synchronization-based rebalancing, from four times to just once in our example.
But, this reduction, whose extent is controlled by the parameter $width$, comes at the expense of potentially expanding unnecessary vertices, and hence {\bf leads to redundant computation}.
For example, if the global queue's length limit $L$ is set to 5, vertices \ding{201} would be pruned following the first parallel epoch.
But with a $width$ of 3 instead of 2, vertices \ding{195} and \ding{201} would be expanded before synchronization.
Consequently, expanding \ding{201} becomes a redundant task, highlighting the inefficiency introduced by an excessively large $width$.

\section{Intuition and Design Choices}
\label{sec:insight}

\subsection{Limitations of Handling Fork-Join Jobs on CPUs}

\begin{figure*}[ht]
  \centering
  \includegraphics[width=\linewidth]{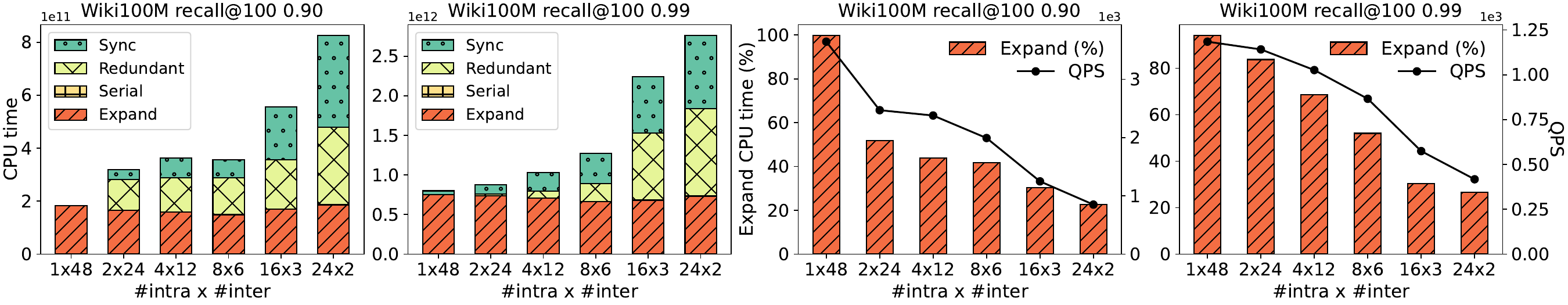}
  \caption{iQAN CPU time breakdown for a single query with varying accuracy and parallelism strategies. ``Serial time'' for the joining phase; ``Expand time'' for useful expansions during the forking phase; ``Redundant time'' for unnecessary expansions; and ``Sync time'' resulting from thread synchronization and load imbalance.}
  \label{fig:CPU-time-acc}
\end{figure*}


While path-wise parallelism reduces synchronization frequency by extending the parallel phase, it still relies on the traditional fork-join model. 
In ANNS, distance calculation tasks during expansion are typically managed by parallel threads spawned in the fork phase.
In contrast, pruning and rebalancing tasks require a comprehensive view of all sub-queues to achieve global ordering, making them suited for the join phase of the fork-join model. 
However, our analysis reveals a fundamental dilemma in using the fork-join model for graph-based ANNS, leading to scalability issues.


Specifically, as we enhance intra-query parallelism by increasing the number of intra-query threads through iQAN to reduce latency, we encounter a significant drop in throughput, measured in Queries Per Second (QPS). 
As demonstrated by Figure \ref{fig:CPU-time-acc}, the throughput decreases by 37.7\%-76.7\% times when intra-query parallelism is scaled from 1 to 24.


To delve into the root cause of this issue, we analyze the CPU time allocated to each query by dividing it into four main categories.
The time dedicated to the joining phase is termed ``serial'' time, solely managed by the master thread. 
Conversely, the parallel forking phase's time is dissected into several components:
1) ``Expand'' time, representing the duration of necessary vertex expansions and is the only period contributing productively to actual progress;
2) ``Redundant'' time, encompassing the unnecessary expansions that would have been pruned in a serial execution, representing an additional tax due to parallelism.
and 3) ``Sync'' time that encapsulates the remaining CPU time outside the forking and joining phases, typically arising from thread synchronization and load imbalance during transitions between these phases.


As indicated in Figure \ref{fig:CPU-time-acc}, a detailed breakdown analysis reveals that the decrease in throughput is mainly attributed to the rise in ``redundant'' and ``sync'' times.
Consequently, the share of CPU time allocated to ``expand'', the only productive phase, is significantly reduced.

\begin{figure*}[ht]
  \begin{minipage}[t]{0.28\linewidth}
    \centering
    \includegraphics[width=\linewidth]{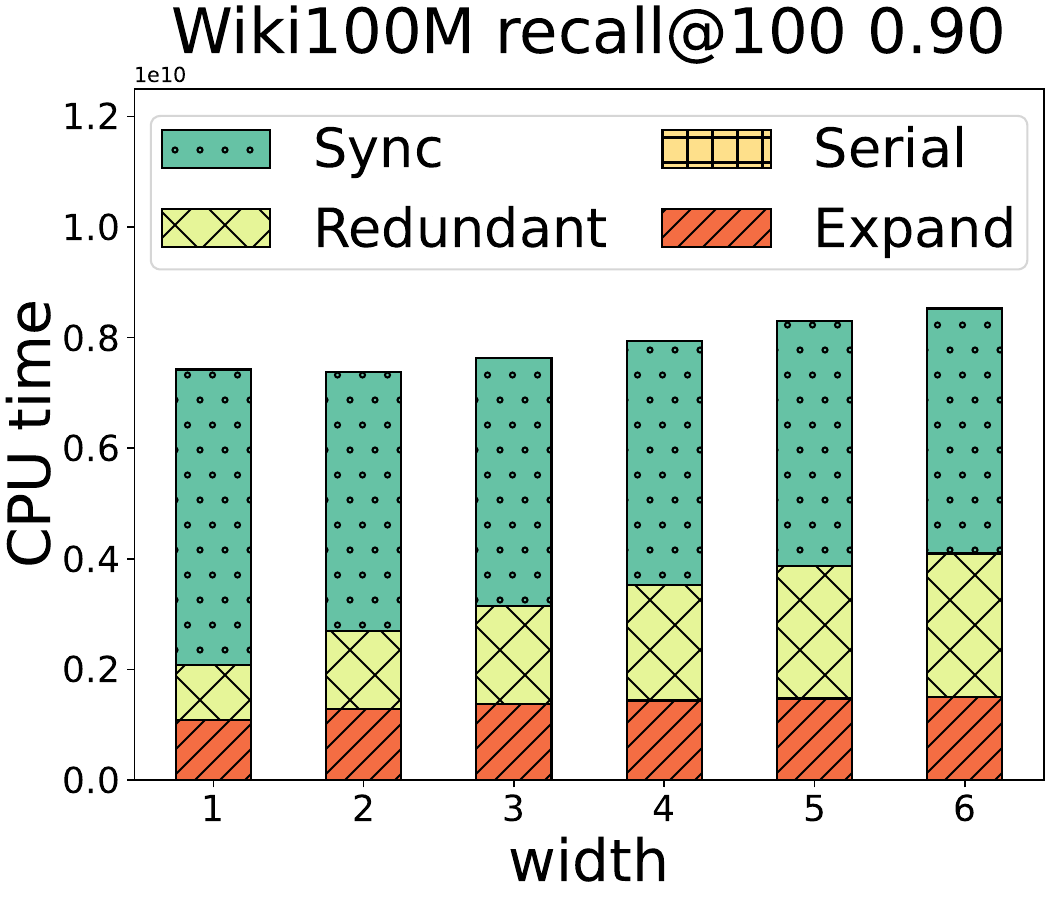}
    \caption{
      iQAN CPU time breakdown for a single query across varying $width$ with 32 threads.}
    \label{fig:tradeoff}
  \end{minipage}
  \hspace{0.01\linewidth}
  \begin{minipage}[t]{0.32\linewidth}
    \centering
    \includegraphics[width=\linewidth]{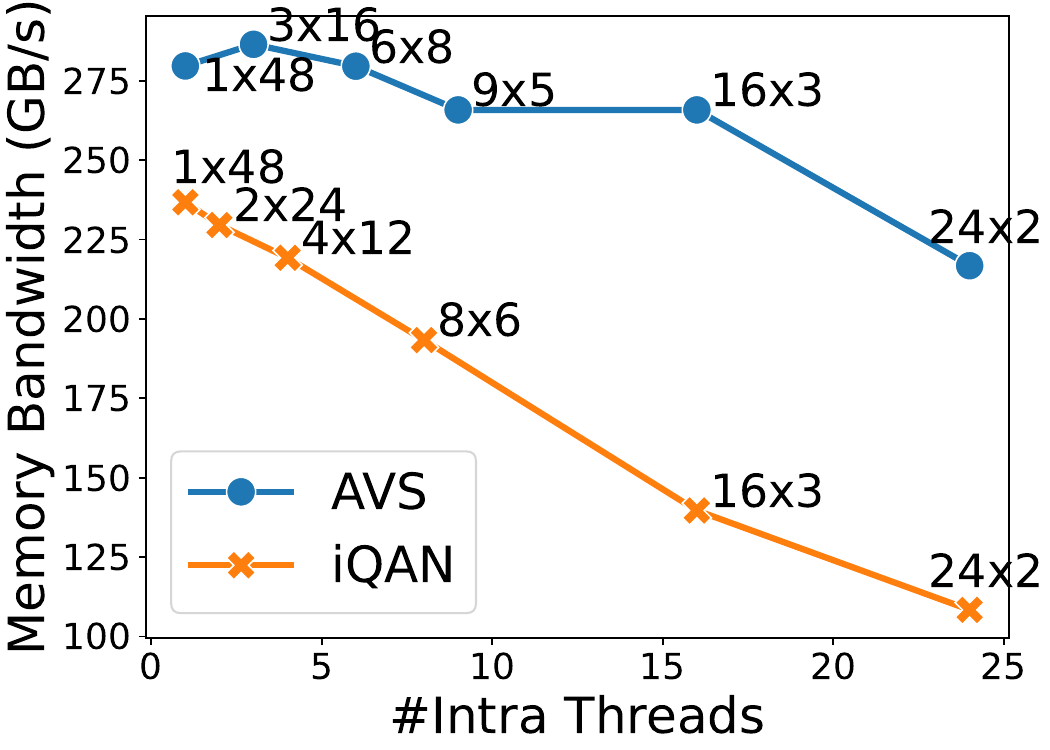}
    \caption{
      Memory bandwidth utilization for different parallelism strategies denoted as ``$intra \times inter$''.}
    \label{fig:iqan-mem-bw}
  \end{minipage}
  \hspace{0.01\linewidth}
  \begin{minipage}[t]{0.31\linewidth}
    \centering
    \includegraphics[width=\linewidth]{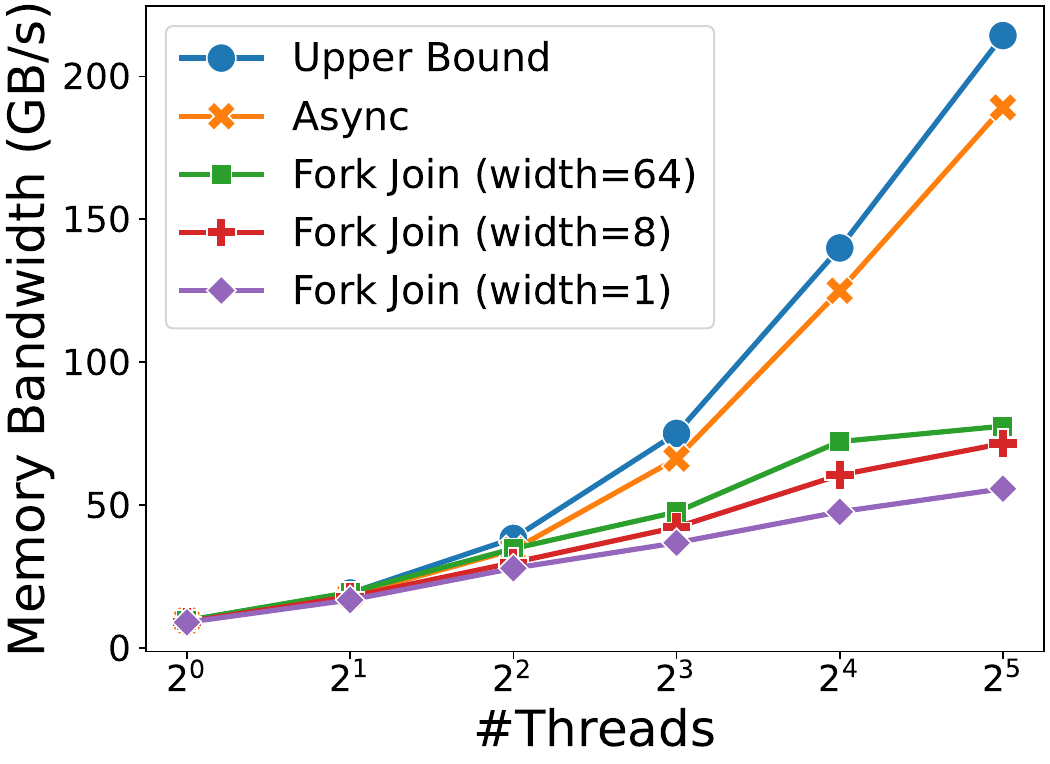}
    \caption{
      Comparison of memory bandwidth utilization in a micro-benchmark performing distance calculations.}
    \label{fig:micro-bench-mem-bw}
  \end{minipage}
\end{figure*}

Moreover, further examination of adjusting the \textit{width} parameter, which controls the size of the parallel section, uncovers a dilemma.
As illustrated in Figure \ref{fig:tradeoff}, a larger parallel section could diminish the synchronization frequency but not the associated overhead.
This is because a decreased synchronization frequency lessens the chances for pruning and rebalancing in the join phase, which results in more redundant calculations and worsens load imbalance.
This reveals handling fork-join tasks on CPUs faces the inherent difficulties in addressing challenges related to synchronization overhead and redundant calculations.


\subsection{Towards a Fully Asynchronous Architecture}
\label{sec:towards-fully-async}

\newcommand{\approptoinn}[2]{\mathrel{\vcenter{
  \offinterlineskip\halign{\hfil$##$\cr
    #1\propto\cr\noalign{\kern2pt}#1\sim\cr\noalign{\kern-2pt}}}}}

\newcommand{\appropto}{\mathpalette\approptoinn\relax}

As the dimensions of embedding vectors increase, ANNS evolves into an extremely memory-intensive application.
In this context, the key to enhancing both throughput and latency lies in optimizing memory bandwidth utilization during the edge expansion phase outlined in Algorithm \ref{alg:expand}. 
Building on our findings, we propose an empirical formula to steer future throughput improvements:
\[ Throughput \appropto EMB = PMB \times (1 - RR) \]
This equation suggests that the overall query throughput is approximately proportional to what we term as ``Effective'' Memory Bandwidth (EMB). 
This metric is calculated by multiplying the {\bf physical memory bandwidth (PMB)} by the factor (1 - RR), where RR represents the {\bf Redundant Ratio}. 
The Redundant Ratio indicates the percentage of vertices that are unnecessarily processed and could have been pruned in a serial execution. 
In the following discussion, we examine how the traditional fork-join model restricts both PMB and RR, and explore the potential benefits of adopting a {\bf fully asynchronous architecture} to overcome these constraints.

\subsubsection{Improving PMB}

Figure \ref{fig:iqan-mem-bw} demonstrates \systemname and iQAN's physical memory bandwidth utilization on the SE-Cohere dataset under various parallelism strategies, measured by hardware counters.
The configuration ``$intra \times inter$'' denotes the distribution of 48 threads across $inter$ concurrent queries, with each processed by $intra$ threads.
Despite using the same total number of threads, increasing intra-query parallelism reduces PMB significantly. 
At a parallelism setting of ``$24 \times 2$'', PMB drops to just 108.39 GB/s, under 40\% of the 266 GB/s maximum bandwidth measured by Intel mlc~\cite{intel-mlc}.

To explore this issue, we designed a micro-benchmark for parallel Euclidean distance calculations.
This benchmark processes calculation between 1024-dimensional fp32 vectors using AVX512.
Within the conventional fork-join paradigm with Intel OpenMP~\cite{dagum1998openmp}, each thread performs a fixed task of $32 \times width$ distance calculations per parallel phase, mimicking the expansion of a vertex with 32 out-edges.
Our findings, depicted in Figure \ref{fig:micro-bench-mem-bw}, indicate that the inefficiencies of fork-join models persist even with significant $width$ settings, with a 32-thread fork-join configuration at a $width$ of $64$ achieving only 36.19\% of the theoretical memory bandwidth.
An asynchronous model, where threads continuously process vertices {\bf without waiting for new tasks to be assigned}, is essential to eliminate bandwidth waste.

\subsubsection{Reducing RR}

Besides the idle time caused by synchronization barriers, redundant computations contribute to another issue: the wasteful use of memory bandwidth.
These unnecessary computations arise from exploring areas with minimal potential.
In iQAN's fork-join model with static distribution, each thread is assigned a fixed number of $width$ vertices, without considering their processing speed. 
As $width$ increases, the likelihood of redundant computations grows due to less frequent global rebalancing.
We expect dynamic workload distribution through an asynchronous framework, where faster threads process more vertices than slower ones.
Although achieving such a perfect balance poses a challenge, {\bf transitioning to an adaptive distribution strategy} could significantly lower the Redundant Ratio, thereby improving both efficiency and resource utilization.

\section{\systemname Design}
\label{sec:arch}

As previously discussed, the key to enhancing ANNS throughput amidst high intra-query parallelism (i.e., achieving low latency)  lies in crafting an architecture that:
1) allows memory-intensive distance calculation threads to continuously process vertices without delays for new tasks, thereby optimizing PMB; 
2) implements an adaptive distribution strategy that dynamically allocates workloads to threads based on their current paces, thus minimizing RR.
And to this end, we introduce a fully asynchronous architecture that deconstructs the traditional BFiS algorithm into three different kinds of components that communicate asynchronously.

Specifically, to enable better parallelization, as depicted in Figure \ref{fig:arch}, we divide the intra- threads handling the same query into multiple sub-groups, which also partitions the original global priority queue, $Q$, into individual sub-queues for each thread sub-group.
Each subgroup comprises a single {\em sub-queue maintainer} (sub-que) thread and multiple {\em distance calculator} (dis-cal) threads. 
In the illustrated case, the thread group handling a single query is divided into two sub-groups, each with one sub-que thread and two dis-cal threads.
The dis-cal threads receive source vertices as tasks, process their outgoing edges, and then the distances for destination vertices are sent back to the sub-que thread, which sorts and merges these results into the sub-queue. 
This setup guarantees that only dis-cal threads perform the memory-intensive Expand operation outlined in Algorithm \ref{alg:expand}.
Ideally, we want the sub-queue never empty so that dis-cal threads will never stall to improve the PMB.
Moreover, to tackle the necessity for global rebalancing and to reduce RR, we introduce a {\em global balancer} thread. 
This thread scans all sub-queues to conduct redistribution and pruning actions based on a comprehensive view. 
Ideally, this redistribution activity should happen as frequently as possible to lower RR effectively.

In this section, we will first use a synchronous design to illustrate how these components cooperate with each other.
Then, we discuss how to disentangle the interactions between the global balancer and sub-queue maintainers, the sub-queue maintainer and distance calculators, respectively, ultimately achieving a fully asynchronous architecture.

\subsection{Straw-man Synchronous Design}
\label{sec:ssd}

\begin{figure}[b]
  \centering
  \includegraphics[width=0.95\linewidth]{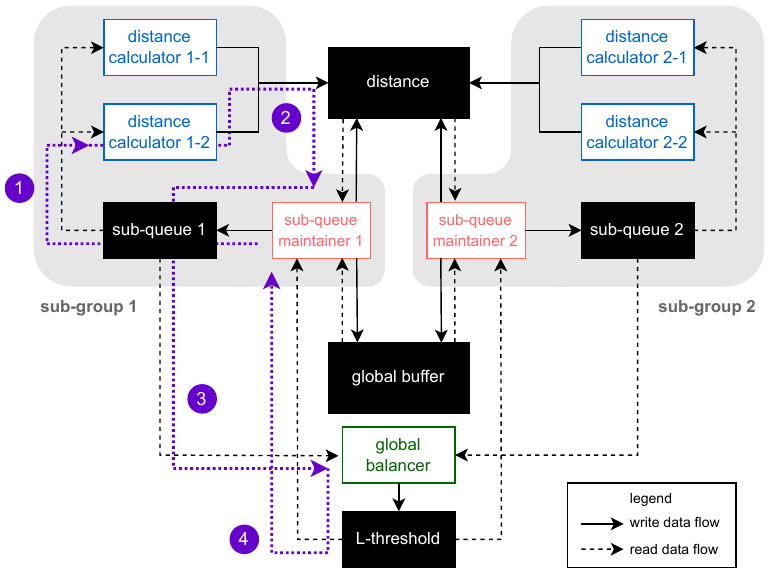}
  \caption{
    An illustration of two thread groups in \systemname.
    Threads (represented by white boxes) interact with each other through shared data structures (represented by black boxes) asynchronously.
    (\S\ref{sec:arch})}
  \label{fig:arch}
\end{figure}

To execute the architecture depicted in Figure \ref{fig:arch}, an initial straightforward synchronous design introduces four types of inter-component dependencies (marked as \ding{192}, \ding{193}, \ding{194}, and \ding{195} in Figure \ref{fig:arch}). 
Before the search, the two sub-queues are initialized with vertices in the entry node set of the graph.
As an illustration, following the edge-wise parallelism strategy, each iteration begins with (\ding{192}) the sub-que thread dispatching the out-edges of the first unchecked vertex in the sub-queue to each dis-cal thread.
After parallel expansion, (\ding{193}) multiple destination vertices are returned to their respective sub-que thread for sorting.
Then (\ding{194}) the global balancer gathers all vertices in the sub-queues, calculates the global order, keeps the top $L$ vertices -- according to the preset global threshold -- and prunes the others.
Finally, (\ding{195}) the top $L$ vertices are redistributed to each sub-queue, and a new iteration starts.

\subsection{Disentangling Global Balancer and Sub-que}
\label{sec:distangle-global-balancer}

The above straw-man implementation faces considerable delays as tasks idly wait for the preceding ones to finish. 
To improve performance, it's crucial to enable tasks to proceed concurrently with less dependency on their predecessors' results. 
This involves employing shared data structures that allow for speculative execution without the latest results. 

\noindent{\bf Queue Pruning (disentangle dependency \ding{194}).}
Decoupling the global balancer from sub-queue maintainers is achieved by maintaining a global variable, the ``L-threshold''.
This entails pruning less promising candidates and guiding sub-queue maintainers towards more potential-rich areas. 
Essentially, this process involves identifying the L-th closest visited vertices according to the global order. 
Yet, accurate L-threshold calculation and sub-queue rebalancing become complex due to concurrent ongoing updates. 
Therefore, we devise an algorithm to {\bf approximate a slightly larger L-threshold} based on asynchronous sub-queue snapshots.
Specifically, the global balancer uses an array of pointers, one for each sub-queue. 
These pointers begin at the ends of the sub-queues and advance forward periodically, 
mimicking a merge-sort operation by moving the pointer with the largest distance forward.
The process continues until only $L$ candidates remain across all sub-queues after simulating pruning extra candidates by moving the pointers, 
at which point the largest element among those before the pointers is identified and announced as the L-threshold.

\noindent{\bf Dynamic Load Balancing (disentangle dependency \ding{195}).}
Given the premise that sub-queue maintainers routinely adjust their sub-queues based on the \textit{L-threshold}, it's implied that the candidates remaining are deemed suitable for future expansion and hence a reduced RR.
However, since the pruning process is governed by distance values rather than queue length, disparities can arise, leading to significant load imbalances where some queues may become substantially longer than others.
To address this issue, we introduce an additional dynamic load-balancing strategy based on work-stealing.
This approach is to enable sub-queue maintainers with shorter queues to assist those with longer ones.
Sub-queue maintainers periodically check if their queue is among the longest.
If so, they help alleviate congestion by transferring some of their queue's elements to the shared buffer.

\subsection{Disentangling Sub-que and Dis-cal}

The essence of separating sub-queue maintainer threads from distance calculator threads in our architecture lies in ensuring that distance calculator threads do not wait for new tasks from the sub-queue maintainers nor directly return the calculated distance results to them.
To navigate around dependency \ding{193}, we establish a distance array for each query.
Each array element, corresponding to a vertex, stores the calculated distance from the query point to this vertex, along with a `ready' flag initially set to 0.
Thus,  distance calculators simply save the calculated results in the array and mark the `ready' flag as 1, allowing sub-queue maintainers to asynchronously retrieve the results from the array later.

Regarding dependency \ding{193}, if a distance calculator completes its assigned edges in the current iteration, it {\bf doesn't idly wait} for the sub-queue maintainer to finish sorting. 
Instead, it preemptively fetches the second unchecked vertex from the sub-queue and begins speculative expansion. 
This speculative process can extend to the third and beyond, but at each speculative execution's beginning, the distance calculator first checks if the current iteration number remains unchanged. 
If it detects an update, indicating the sub-queue maintainer has completed a sorting and pruning round, the distance calculator resets and starts processing from the first unchecked vertex again.

This method enables adaptive task distribution without the need for a centralized coordinator. 
Only dis-cal threads that operate more quickly in a given cycle will process additional speculative vertices, 
thus minimizing redundancy in comparison to the fixed-width strategy utilized by iQAN. 
We adopt edge-level rather than vertex-level or path-level parallelism to facilitate a more evenly distributed workload when paired with our strategy's ability to balance workloads dynamically.

Another inefficiency in existing methods is the redundant calculation of distances for destination vertices pointed to by multiple source vertices. 
However, by maintaining a distance array and `ready' flags, distance calculators can check and skip unnecessary computations. 
While this skipping could complicate workload balancing in static parallelism settings, our adaptive strategy naturally mitigates this issue.


\begin{figure}[ht]
  \centering
  \includegraphics[width=\linewidth]{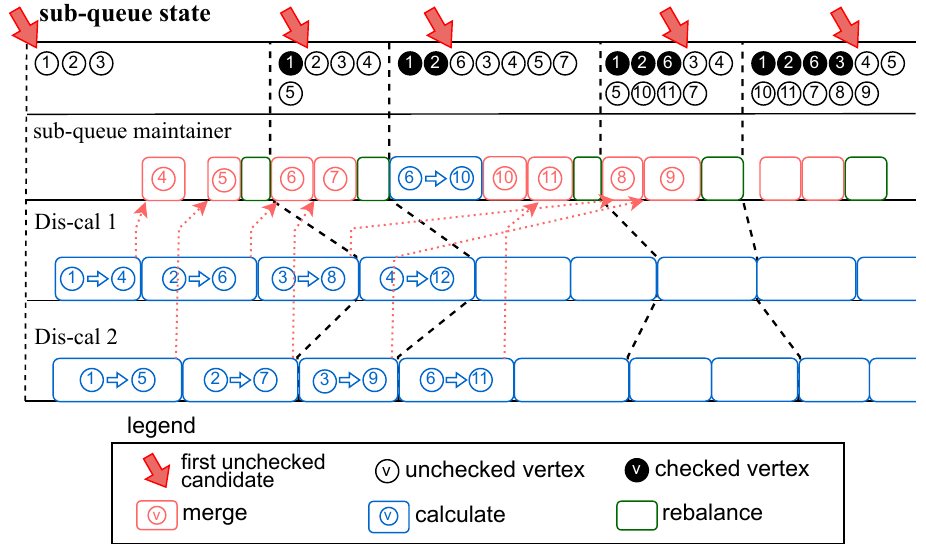}
  \caption{
    A timeline depicting the execution of a sub-queue maintainer thread and two distance calculator threads searching in the example graph of Figure \ref{fig:example-graph-0}.}
  \label{fig:async-model}
\end{figure}

To further demonstrate our strategy, Figure \ref{fig:async-model} depicts how a sub-que thread and two dis-cal threads operate within the example graph shown in Figure \ref{fig:example-graph-0}. 
The dashed lines in the figure denote misaligned asynchronous epochs, triggered by updates to the first unchecked candidate in the sub-queue, highlighted by red arrows in the figure.
After the expansion of each edge,
the dis-cal threads periodically check whether the first unchecked candidate in the sub-queue has been updated, guiding their actions for each epoch. 

As we can see from the example, initially, the distance calculators expand vertex \ding{192}, the first unchecked candidate. 
Each distance calculator processes an outgoing edge from \ding{192}, in an edge-wise approach. 
Subsequently, as the first unchecked vertex remains unchanged, they speculatively advance to the next unchecked candidates, vertices \ding{193} and \ding{194}.

In this initial epoch, Dis-cal 1 speculatively processes the neighbors of \ding{193} ($\Rightarrow$\ding{197}) and \ding{194} ($\Rightarrow$\ding{199}), while Dis-cal 2 handles \ding{193}'s other neighbor ($\Rightarrow$\ding{198}). 
Concurrently, the sub-queue maintainer retrieves distances for vertices \ding{195} and \ding{196} from the distance array, merges them into the sub-queue, and updates the status of vertex \ding{192} to  ``checked'', thereby marking the start of a new epoch.

As the expansion progresses, upon completing their tasks for vertices \ding{199} and \ding{198}, Dis-cal 1 and 2 both notice that the first unchecked candidate has shifted from vertex \ding{192} to \ding{193}.
But, since \ding{193} has already been speculatively expanded, they continue speculative processing of the out-edges for \ding{194} and \ding{195}.
During this epoch, the sub-queue maintainer effortlessly retrieves distances for \ding{193}'s neighbors from the distance array without any delay, benefiting from the speculative calculations done in the prior epoch.

In the subsequent third epoch, vertex \ding{197} emerges as the new first unchecked vertex. 
Upon encountering \ding{197}, which neither distance calculators had previously speculatively expanded, they prioritize \ding{197}'s expansion rather than continue their speculative execution. 
From the sub-queue maintainer's perspective, since the distance for \ding{197}'s neighbor, vertex \ding{201}, is unavailable, it calculates this distance and updates the \verb|distance| array accordingly. 
This efficient cycle of speculative processing and asynchronous updating exemplifies the dynamic and continuous workflow enabled by \systemname, which minimizes latency and optimizes throughput in real-time ANNS applications.


\subsection{Implementation Details of \systemname}

To integrate the concepts discussed earlier, this section presents the logic of each component within the thread group. 
Each thread follows the workflow illustrated in Figure \ref{fig:arch}, continuously polling the upstream data structure and writing to the downstream data structure. 
Algorithms \ref{alg:global-balancer-code}, \ref{alg:subq-maintainer-code}, and \ref{alg:dis-cal-code} provide the pseudocode for the global balancer, sub-queue maintainer, and distance calculator, respectively.

The global balancer is responsible for calculating the $L$-threshold. While its functionality can be inlined into the sub-queue maintainers when the number of threads in each group is relatively small, a dedicated thread is assigned to handle this task when low latency is required and sufficient threads are available. This approach ensures frequent pruning.

The sub-queue maintainers are tasked with expanding candidates in their local sub-queues. 
They first attempt to read distances from the distance array to avoid redundant computation.
After each expansion, they prune their sub-queues based on the $L$-threshold and perform load balancing via the global buffer.

The distance calculator threads compute the out-edges of unchecked candidates in the sub-queue in an edge-wise manner, storing their results in the distance array. They also speculatively process other unchecked vertices in the sub-queue if the queue is not updated in time.

\renewcommand{\algorithmicrequire}{\textbf{Requires:}}
\renewcommand{\algorithmicensure}{\textbf{Modifies:}}

\begin{algorithm}
  \begin{algorithmic}[1]
    \Require {Sub-queues $Q[\;]$, the length limit $L$}
    \Ensure {The L-threshold $L-thresh$}
    \While {the search has not ended}
      \State {$L-thresh \gets$ the $L$-th nearest distance among all elements in $Q[\;]$}
    \EndWhile
  \end{algorithmic}
  \caption{The Global Balancer}
  \label{alg:global-balancer-code}
\end{algorithm}

\begin{algorithm}
  \begin{algorithmic}[1]
    \Require {Query $q$, graph $G$, distance array $D[\,]$, distance function $\delta$, L-threshold $L-thresh$, global buffer $B$}
    \Ensure {Sub-queue $Q$, distance array $D[\,]$, global buffer $B$}
    \While {there's unchecked vertice in $Q$}
      \State {$v \gets$ the first unchecked candidate in $Q$}
      \State {$v.state \gets checked$}
      \ForAll {neighbor $u$ of $v$ in $G$}
        \If {$u$ is not visited}
          \If {$D[u]$ is not set}
            \State {$D[u] \gets \delta(u, q)$}
          \EndIf
          \State {mark $u$ as visited}
          \State {mark $u$'s state as unchecked}
          \State {$Q \gets Q \cup \{u\}$}
        \EndIf
      \EndFor
      \State {prune candidates larger than $L-thresh$ in $Q$}
      \If {$Q$ is among the longest sub-queues}
        \State {offload some candidates into the global buffer $B$}
      \Else
        \State {pick some candidates from $B$ into $Q$}
      \EndIf
    \EndWhile
  \end{algorithmic}
  \caption{The Sub-queue Maintainer}
  \label{alg:subq-maintainer-code}
\end{algorithm}

\begin{algorithm}
  \begin{algorithmic}[1]
    \Require {Query $q$, graph $G$, sub-queue $Q$, distance function $\delta$, distance array $D[\;]$}
    \Ensure {Distance array $D[\;]$}
    \While {there's unchecked vertice in $Q$}
      \State {$v \gets$ the first unchecked candidate in $Q$}
      \State {$v' \gets v$}
      \While {$Q$'s first unchecked candidate is $v'$}
        \ForAll {neighbor $u$ of $v$ in $G$}
          \If {$u$ is not visited \& $D[u]$ is not set}
            \State {$D[u] \gets \delta(u, q)$}
          \EndIf
        \EndFor
        \State {$v \gets$ the next unchecked candidate in $Q$}
      \EndWhile
    \EndWhile
  \end{algorithmic}
  \caption{The Distance Calculator}
  \label{alg:dis-cal-code}
\end{algorithm}





\section{Evaluation}

\subsection{Evaluation Methodology}

\noindent \textbf{Datasets.}
Our evaluation includes six datasets of float vectors.
In addition to the traditional benchmark datasets SIFT100M \cite{sift1b} and DEEP100M \cite{deep1b}, which include 100 million 128-dimensional and 96-dimensional float vectors,
we also evaluate \systemname with three more recent embedding datasets derived from large embedding models.
SE-OpenAI~\cite{wikipedia-stephanst} and SE-Cohere~\cite{wikipedia-cohere-2022} comprise 400 thousand embedding vectors derived from Simple English Wikipedia~\cite{wikipedia-simple}.
These vectors are encoded using the 1536-dimensional OpenAI text-embedding-3-small embedding model~\cite{openai2024text-embedding-3-small} and the 768-dimensional Cohere EmbedV3 model~\cite{cohere2022multilingual}, respectively.
Additionally, we use WIKI100M, which includes 100 million embedding vectors from a random subset of the entire Wikipedia site, processed using two Cohere models that produce embedding vectors with 768~\cite{wikipedia-cohere-2022} and 1024~\cite{wikipedia-cohere-2023} dimension, respectively.

\noindent \textbf{Comparison.}
We have implemented \systemname in over 7000 lines of C++ code. 
Our implementations support multiple similarity graphs, including NSG \cite{fu2017fast}, SSG \cite{fu2021high}, and Vamana index \cite{jayaram2019diskann} constructed using ParlayANN \cite{manohar2024parlayann}.
For each dataset, we evaluate the search performance using the most accurate index that can be constructed within a reasonable time.
We use NSG for relatively small datasets, and Vamana index for Wiki100M, since their construction time of NSG exceeds 7 days.
Also, we integrate \systemname into FAISS~\cite{faiss-code}, making intra-query parallelism available within FAISS' framework.

We compare \systemname and FAISS-\systemname with iQAN~\cite{peng2023iqan}, the state-of-the-art parallel ANNS system, ParlayANN~\cite{manohar2024parlayann}, the state-of-the-art index construction engine which also support searching, and Milvus~\cite{wang2021milvus}, FAISS~\cite{douze2024faiss}, two leading vector search engines. 
For iQAN and ParlayANN, search hyper-parameters are adopted as suggested by the provided script.

To evaluate the performance of the tested systems, we measure latency and Queries Per Second (QPS) while targeting a predefined accuracy threshold.
Accuracy is gauged using recall@100, which verifies the accuracy of the search results by confirming the presence of the true top 100 nearest neighbors.
Each test fully utilizes all 48 physical cores of our testing socket and experiments with various parallelism strategies, both inter-query and intra-query.
Our accuracy benchmarks range from 0.9 to 0.995, covering a broad spectrum of vector search applications.

\noindent \textbf{Test Platform.}
Our experiments are conducted on a system equipped with an Intel\textregistered\; Xeon\textregistered\; Platinum 8457C processor (2.60GHz), featuring 48 physical cores of a socket and 528 GB of DDR5 DRAM. 
The maximum physical memory bandwidth, measured by mlc~\cite{intel-mlc}, is 266 GB/s.

\begin{figure*}[ht]
  \centering
  \includegraphics[width=0.95\linewidth]{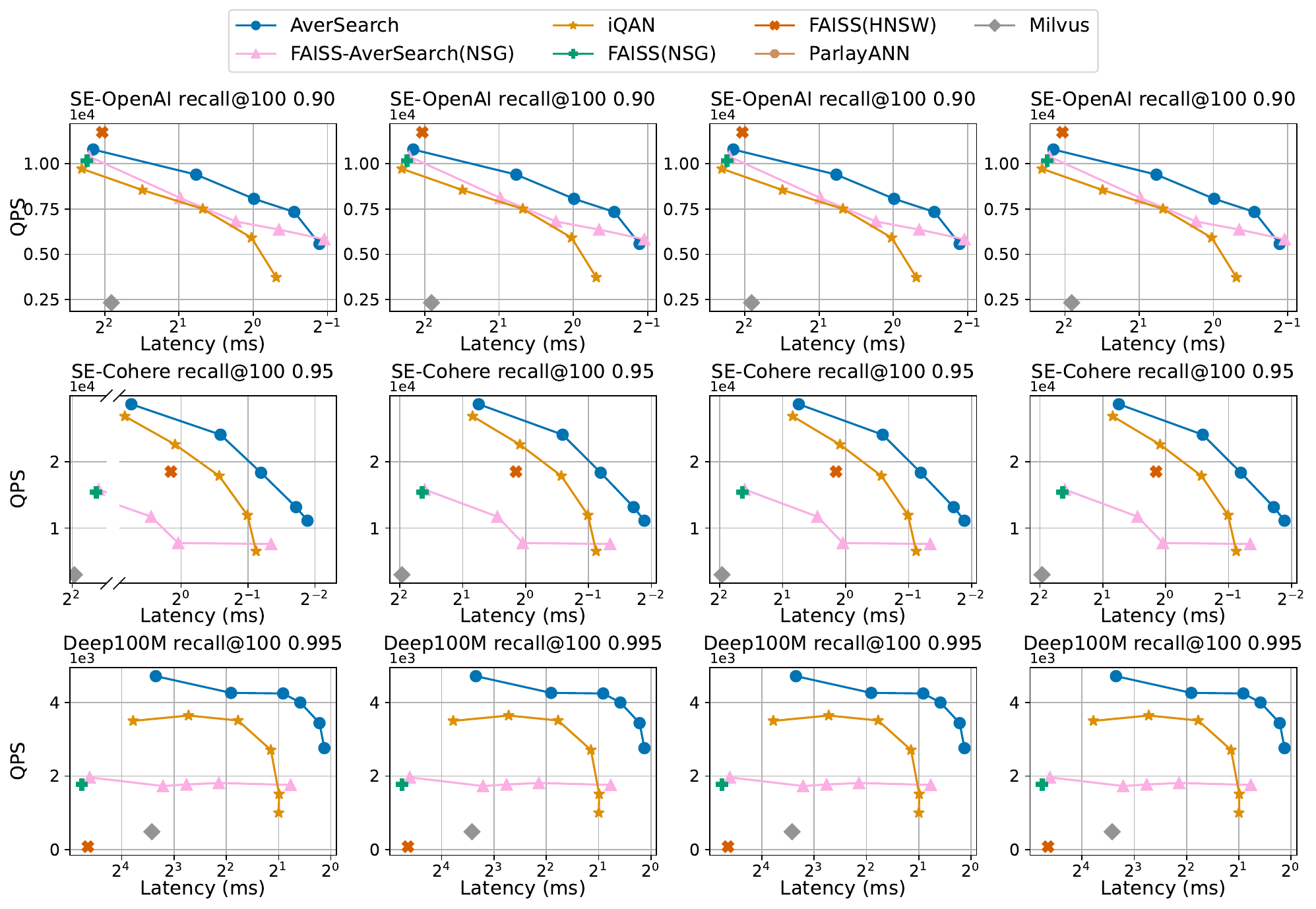}
  \caption{QPS-latency curve under various accuracy targets in different embeddings.}
  \label{fig:eval-simple}
\end{figure*}

\subsection{Evaluation of the QPS-Latency Curve}

%
%
%

We illustrate the trade-off between QPS and latency for \systemname, FAISS-\systemname and iQAN in Figures \ref{fig:eval-deep-sift} and \ref{fig:eval-simple}, where QPS is plotted on the y-axis and latency is on the x-axis in reverse order using a logarithmic scale (base 2).
For \systemname, FAISS-\systemname, and iQAN (i.e., systems with configurable parallelism settings), we adjust the number of intra- and inter-threads to generate the curve.
In all configurations, the total thread count (i.e., intra-threads multiplied by inter-threads) is fixed at 48, matching the total available cores on our machine.
We evaluated only the ``1x48'' configuration for systems without tunable parallelism settings.
For systems with tunable settings, we evaluated configurations including ``1x48'', ``2x24'', ``3x16'', ``4x12'', ``6x8'', ``8x6'', ``12x4'', ``16x3'', and ``24x2'', and plotted the convex hull in Figures 1 and 10.

These figures demonstrate a trade-off between achieving low latency and high throughput, as both \systemname and iQAN experience a decrease in throughput when increasing intra-query parallelism.
However, as evident from the figures, \systemname consistently achieves higher throughput and lower latency than iQAN under the same parallelism settings.

For instance, \systemname's throughput is 1.23-2.38$\times$ higher than iQAN on the DEEP100M and SIFT100M datasets and, {\bf simultaneously}, 1.07 to 2.35 times lower latency when comparing the same parallelism setting with each other.
This performance gap persists in recent datasets featuring larger dimensional embeddings, such as the Wiki100M dataset, where \systemname achieves up to 2.07 times higher throughput.
Moreover, if we compare the QPS achieved at a comparable latency level (less than 20\% diff), \systemname can achieve 2.17$\times$ (SE-OpenAI) to 8.93$\times$ (DEEP100M) higher throughput than iQAN.

The evaluation of FAISS-\systemname and FAISS utilize the FAISS-constructed NSG index, which differs slightly from the original NSG index adopted in our experiment for \systemname and iQAN, as some FAISS-specific parameters are included in the FAISS-constructed index.
To enable a fair comparison, we utilize the same NSG parameters to construct both types of NSG index.
FAISS-\systemname demonstrates satisfactory scalability as the number of intra-threads increases, indicating that \systemname could lower latency without compromising QPS.

We also evaluated the search implementation provided by ParlayANN~\cite{manohar2024parlayann} on Wiki100M and Wiki100M-2, for which the indices are built using it.
Since ParlayANN does not provide intra-query parallelism optimization, it demonstrates comparable performance to \systemname on the ``1x48'' configuration.

The superior performance of \systemname is attributed to its fully asynchronous architecture, which significantly reduces synchronization overhead by eliminating the use of any barriers, thereby also reducing latency.
Additionally, due to its high resource utilization under settings with intense intra-concurrency, \systemname typically achieves lower minimum average latency than iQAN.
Across various datasets, the minimum average latency reduction ranges from 1.54$\times$ (SE-OpenAI) to 1.96$\times$ (DEEP100M), highlighting \systemname's efficiency in real-world scenarios.

\subsection{Performance Analysis}
\label{sec:perf-analysis}

%
%

As discussed in \S\ref{sec:towards-fully-async}, enhancing throughput hinges on improving PMB and reducing the RR. 
Table \ref{tab:perf-analysis-pmb} presents these metrics measured across various datasets with the recall set to the highest level, 0.995. 
For instance, \systemname achieves 1.68$\times$ higher PMB and a 74$\%$ reduction on RR when compared to iQAN on the Wiki100M dataset. 
According to the formula outlined in \S\ref{sec:towards-fully-async}, this leads to an approximately  ``$149*(1-0.08) : 89*(1-0.31) = 2.2\times$'' higher Effective Memory Bandwidth (EMB), 
which results in a 1.78$\times$ higher throughput according to our evaluation.

In summary, \systemname consistently achieves a higher PMB (1.35$\times$ - 2.32$\times$) than iQAN. 
This improvement is primarily due to the asynchronous design of the distance calculator threads in \systemname, which do not need to wait for sub-queue maintainers. 
However, despite this no-wait strategy, the PMB achieved by \systemname does not always reach the maximum physical bandwidth of the machine, especially for datasets with shorter vectors like SIFT100M and DEEP100M. This limitation often arises because the queue runs out of vertices, preventing further speculative execution.  
This challenge is largely due to the static assignment of thread roles, showing that while \systemname offers more flexibility than iQAN, it still faces hurdles in balancing workloads between sub-queue maintainers and distance calculators.

Moreover, there is a notable reduction in RR for the Wiki100M dataset, largely due to our dynamic work dispatching policy. 
However, the RR for \systemname is comparable to that of iQAN on the two Simple English datasets, mainly attributed to the smaller sizes of these datasets.
This comparison underscores the enhanced effectiveness of \systemname with larger datasets.


\begin{table}[h]
  \centering
  \begin{tabular}{|c|c|c|c|c|}
    \hline
    \multirow{2}{*}{Dataset} & \multicolumn{2}{c|}{PMB} & \multicolumn{2}{c|}{RR} \\
    \cline{2-3} \cline{4-5} & Ours & iQAN & Ours & iQAN \\
    \hline
    Sift100M (128D) & 30 ($2.18\times$) & 13 & 0.06 & 0.22 \\
    Deep100M (96D) & 32 ($2.32\times$) & 13 & 0.09 & 0.30 \\
    SE-Cohere (768D) & 190 ($1.91\times$) & 99 & 0.32 & 0.36 \\
    SE-OpenAI (1536D) & 242 ($1.78\times$) & 135 & 0.38 & 0.36 \\
    Wiki100M (768D) & 149 ($1.68\times$) & 89 & 0.08 & 0.31 \\
    Wiki100M-2 (1024D) & 140 ($1.35\times$) & 103 & 0.34 & 0.64 \\
    \hline
  \end{tabular}
  \caption{Physical Memory Bandwidth (PMB) and Redundancy Ratio (RR) of \systemname (ours) and iQAN across various datasets, targeting a recall@100 of 0.995.}
  \label{tab:perf-analysis-pmb}
\end{table}

\subsection{\systemname and Vector Databases}
\label{sec:vd}


Besides iQAN, we also evaluated \systemname against leading vector databases such as Milvus \cite{wang2021milvus, guo2022manu} and FAISS \cite{douze2024faiss}.
For Milvus, we utilize Hierarchical Navigable Small World (HNSW)~\cite{malkov2018efficient} as the indexing method, as it is the most matured graph-based index available in the system.
For FAISS, we evaluate both HNSW and NSG.
We carefully selected the searching parameters to ensure that the $L$ parameter during searches in both Milvus and FAISS matched our system's (actually slightly smaller than our system to ensure a fair comparison), hence achieving comparable levels of recall.

As indicated in Figure \ref{fig:eval-deep-sift} and \ref{fig:eval-simple}, vector databases achieve lower throughput compared to iQAN and \systemname. 
Milvus and FAISS cannot process the large Wiki100M datasets due to OOM.
However, it is important to note that this is not a head-to-head comparison because fully-fledged vector databases often include additional functionalities that might affect their performance.
Moreover, since both Milvus and FAISS support only inter-query parallelism, we assessed their performance using a ``1x48'' configuration.
The results show that \systemname achieves 1.48$\times$ to 9.82$\times$ lower latency and typically also a higher throughput (up to 5.68$\times$) compared to these vector databases at ``24x2'' for high recall.
Therefore, our system offers a compelling alternative that achieves significantly lower latency.

\subsection{\systemname and Quantization-based Methods}

To assess the significance of accelerating search on the graph-based index, 
we also evaluated Product Quantization method provided by FAISS~\cite{douze2024faiss}.
We test the FlatPQ index which compress the database with product quantization codes.
Since the FlatPQ index is unable to exceed the accuracy of 0.90 on small datasets, and 0.70 on larger ones,
we only test it on the two Simple English datasets with a recall target of 0.90.
The 1536-dimensional SE-OpenAI vectors were encoded into 512-length 8-bit codes, and the 768-dimensional SE-Cohere vectors into 256-length 8-bit codes.
FlatPQ index achieves 639.55 and 691.55 QPS on SE-OpenAI and SE-Cohere respectively, which is 0.59 and 0.23 times of \systemname.
Although the ``curse of dimensionality'' limits the accuracy of quantization methods, our system's acceleration can still speed up hybrid methods of quantization- and graph-based methods.

\subsection{Breakdown Analysis}
\label{sec:eval-breakdown}

%
%
%
%

\begin{figure}[h]
  \centering
  \includegraphics[width=0.85\linewidth]{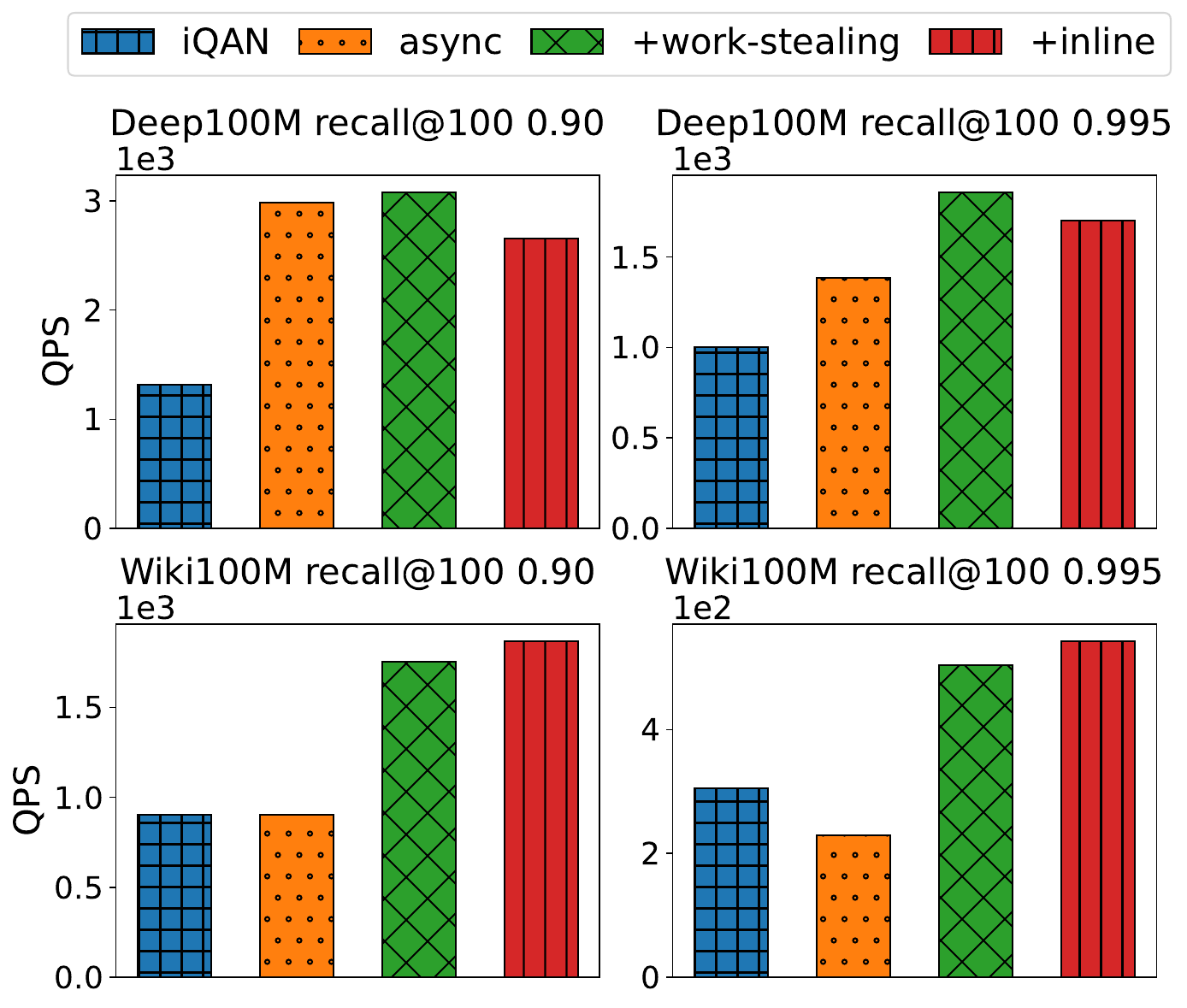}
  \caption{Breakdown Analysis at ``24x2''.}
  \label{fig:perf-breakdown}
\end{figure}

To further validate the optimization sources in \systemname, we present a performance breakdown in Figure \ref{fig:perf-breakdown}. 
In this analysis, the ``async'' refers to the basic implementation of \systemname's fully asynchronous architecture, but without the dynamic work-stealing balance discussed in \S\ref{sec:ssd}.
As we can see from the figure, ``async'' is already 1.38$\times$ to 2.27$\times$ faster than iQAN on smaller datasets like DEEP100M.
However, this basic setup may encounter severe load imbalances. 
Thus, by incorporating the work-stealing technique, we observe significant improvements, with throughput increases ranging from 1.94$\times$ to 2.20$\times$ on Wiki100M (indicated as ``+work-stealing'').

Additionally, our analysis reveals that for larger graphs with high-dimensional vectors, such as Wiki100M, it is beneficial to partially inline the distance computation within the sub-queue maintainers' process. This ``+inline'' adjustment results in a modest increase in QPS, due to increased PMB. However, this optimization does not translate well to smaller datasets like DEEP100M. 
Consequently, we have implemented this feature as an optional setting.

\section{Related Work}

In this paper, we focus on enhancing intra-query parallelism for CPU-based graph ANNS indices. 
Besides the related works discussed in \S\ref{sec:background}, the field has also seen various orthogonal advancements that complement our approach.
For instance, Product Quantization (PQ) techniques~\cite{jegou2010product, jegou2011searching, ge2013optimized, wei2020analyticdb, wu2017multiscale, wang2020deltapq, abdelhadi2019accelerated}, have been widely adopted to achieve dimensionality reduction. 
These techniques can be seamlessly integrated with our approach as a preprocessing step, allowing for a more flexible balance between performance and accuracy.

Moreover, many efforts have been made to offload part or the entirety of the distance calculation process to faster computing accelerators, such as GPUs~\cite{johnson2019billion, zhao2020song, groh2022ggnn, yu2022gpu, ootomo2023cagra} and FPGAs~\cite{zhang2018efficient, dai2017foregraph}.
However, these solutions often require the dataset to fit within the limited memory capacity of the accelerators, which is prohibitively costly for managing large-scale datasets. 
On the other end of the spectrum, some researchers have turned to high-speed SSDs~\cite{liang2022vstore} and the emerging Compute Express Link (CXL) memory~\cite{jang2023cxl} to handle larger datasets.
These approaches usually retain an in-memory graph index search component,
which could benefit from the enhancements presented in our method.


\section{Conclusion}

This paper presents \systemname, an innovative parallel graph-based ANNS framework with a fully asynchronous architecture. It effectively eliminates unnecessary synchronization (improving PMB) and reduces redundant vertex processing (decreasing RR). 
These advancements allow \systemname to successfully balance the competing demands of throughput, accuracy, and latency. Extensive evaluations across multiple datasets show that \systemname achieves up to 8.9$\times$ higher throughput at the same level of latency, and reduces minimum latency by up to 1.9$\times$ compared to leading systems.

\bibliographystyle{ACM-Reference-Format}
\bibliography{refs/main}

\end{document}